\documentstyle[aaspp4]{article}
 
\lefthead{Elmegreen}
\righthead{The Largest Stars}
\slugcomment{{\it Astrophysical Journal}, {\bf 539}, Aug 10, 2000}
\begin{document}
 
\title{Modeling a high mass turn down in the 
stellar initial mass function}

\author{Bruce G.~Elmegreen\altaffilmark{1}} \altaffiltext{1}{IBM
Research Division, T.J. Watson Research Center, P.O. Box 218, Yorktown
Heights, NY 10598, bge@watson.ibm.com}

\begin{abstract} Statistical sampling from the stellar initial mass
function (IMF) for all star-forming regions in the Galaxy would lead
to the prediction of $\sim1000$ M$_\odot$ stars unless there is a rapid
turn-down in the IMF beyond several hundred solar masses.  Such a turndown
is not necessary for dense clusters because the number of stars sampled
is always too small. Although no upper mass limits to star formation
have ever been observed, a theory for the IMF should be able to explain
the lack of $\sim1000$ M$_\odot$ stars in normal galaxy disks.

Here we explore several mechanisms for an upper mass cutoff, including
an exponential decline of the star formation probability after a
turbulent crossing time. The results are in good agreement with the
observed IMF over the entire stellar mass range, and they give a gradual
turn down compared to the Salpeter function above $\sim100$
M$_\odot$ for normal thermal Jeans mass, $M_J$. However, they cannot
give both the observed power-law IMF out to the high-mass sampling limit
in dense clusters, as well as the observed lack of supermassive stars in
whole galaxy disks. The exponential decline is too slow for this. Either
there is a sharp upper mass cutoff in the IMF, perhaps from
self-limitation, or the IMF is different for dense clusters than for the
majority of star formation that occurs at lower density. In the latter
case, dense clusters would have to form an overabundance of massive
stars relative to the average IMF in a galaxy. Evidence for a difference
in the cluster and field IMFs supports this picture, but
systematic effects could mimic this evidence even with
a universal IMF. 

Within the framework of the sampling model, the upper mass turn down
should shift toward higher mass when $M_J$ shifts upward, as might be
the case in some starburst galaxies, and shift toward lower mass when
$M_J$ is lower, as might be the case in ultracold or high pressure
regions. Supermassive stars may therefore be possible in starburst
galaxies, while in low surface brightness regions, where ultracold gas
might exist at normal pressures, or in galactic cluster
cooling flows where cold gas
could have extremely high pressures, a high fraction of the star
formation could end up as brown dwarfs. \end{abstract}

received ... November 10, 1999, revised ... March 8, 2000

\section{Introduction}

A recent model for the stellar initial mass function (IMF), in which
stellar masses are randomly sampled down to the thermal Jeans mass from
hierarchically-structured, pre-stellar clouds (Elmegreen 1997b, 1999a,
2000a; hereafter Papers I, II, and III), is extended here to consider the
simultaneous evolution of small and large turbulent cloud structures,
introducing an additional competition for mass that was not present in
the earlier work. As a result, the slope of the power law comes closer
to the Salpeter IMF ($M^{-1.3}d\log M$ changes to
$M^{-1.35}d\log M$), and there is an interesting turn down in the IMF at
high mass that might have important consequences. Such a turn down has
not yet been observed in real clusters because the masses of the largest
stars generally scale with the total cluster mass, as expected for
random sampling (Elmegreen 1983, 1997b; Schroeder \& Comins 1988; Massey
\& Hunter 1998; Selman et al. 1999), but statistical considerations
(Sect. II) suggest there should be a turn down in the IMF where this
scaling stops, somewhere in the range from 100 to 1000 M$_\odot$.
Otherwise, the largest star in a galaxy like ours would be several
thousand solar masses, and it would be even larger in starbursts.

There have been several previous suggestions that stars have a maximum
mass (Larson 1982; Myers \& Fuller 1993; Khersonsky 1997). Larson \&
Starrfield (1971), Kahn (1974), and Wolfire \& Cassinelli (1986, 1987)
suggested that radiation pressure on dust grains halts the accretion
onto a massive star, thereby limiting the mass. However, there is no
obvious dependence between maximum stellar mass in a cluster and
metallicity (Freedman 1985; Massey, Johnson, \& DeGioia-Eastwood 1995b),
so radiation pressure effects are probably not important. Besides, a
large part of the stellar accretion may be optically thick through an
accretion disk (e.g., Nakano 1989, Nakano, Hasegawa, \& Norman 1995;
Jijina \& Adams 1996) or protostar coalescence (see Sect.
\ref{sect:coal}), and in that case, radiation pressure effects would not
be expected.

There seems to be no fundamental limit to the stellar mass {\it at the
time of birth} from the Eddington limit. The most massive stars have
their luminosities increase roughly proportional to mass (Massey 1998), so
they are probably near their Eddington limits on the main sequence and
consequently shed a lot of mass after they form. Such rapid mass loss is
observed around the most massive stars (Figer et al. 1998; Morse et al.
1998; Grosdidier et al. 1998). However, there would seem to be nothing
to prevent the rapid accumulation of optically thick gas {\it during}
star formation, especially if it comes in through a dense disk or by
some pre-stellar coalescence. As long as the accretion rate during star
formation exceeds the maximum wind loss rate, which is about the stellar
density at one optical depth on the surface multiplied by the surface
area and the escape velocity, and as long as the total accretion time is
less than the nuclear burning time in the core, which is about
$2\times10^6$ years for a massive star, then a star with an arbitrarily
large mass can form. 

A more important question is whether other things happen first in a
massive clump that makes most of the gas unavailable for accretion to a
single star. Here we examine the possibility that star formation in the
smaller, denser, and faster-evolving subclumps is likely to happen
first, and that competition for gas and cloud dispersal, 
rather than self-destruction, are
what limit the {\it initial} mass of the largest star that can form. 
This solution is obvious in the extreme case: whole molecular
clouds do not form single supermassive stars because smaller stars
form first and either take the gas themselves or disperse it away 
from the cloud.  The same competition for mass should arise on smaller
scales too, preventing the formation of $\sim1000$ M$_\odot$
stars, for example, in $10^4$ M$_\odot$ cores,
by the earlier formation of $1-100$ M$_\odot$ stars in the same gas. 

Some level of 
competition between low-mass star formation in small clumps and more
massive star formation in the larger clumps that contain them was
already present in the models of Papers I--III. It had the effect of
steepening the mass function from $M^{-1.15}d\log M$ for pure
root-density weighted sampling to $\sim M^{-1.3}d\log M$ after numerical
simulations of this process. But there was no {\it constraint on timing}
for this type of mass competition as there is in the present models.
This means that the Salpeter function would extend to arbitrarily high
masses if a large clump can wait forever to form its star. However, such
infinite waiting times are unrealistic because the gas inside a large
clump should continuously move and make new small clumps, forming
additional small stars, and because the winds and radiation from stars
that have already formed in the small clumps should significantly
alter the conditions for future star formation in the large clump. If
the gas that was originally available for the large star is
systematically channeled into small stars first, or if this gas is removed
from the large-scale clump altogether, then a large star will not
form there. 

We believe this is a realistic situation for turbulent clouds and
attempt to model it here within the context of the random sampling model
for the IMF. We do this by including an exponentially decreasing
function, $e^{-\omega t(M)/t(M_J)}$, for the probability of forming a star on a
certain scale; $t(M)$ is the crossing time on the scale that forms a
star of mass $M$, and $M_J$ is the mass at the lower limit to the power
law in the IMF, considered here to be the thermal Jeans mass; $\omega$
is a constant of order unity. With this
function, large clumps with long crossing times get a decreased chance
of forming a massive star. 

In what follows, we first present evidence that there should be a turn
down in the IMF for normal star-formation regions at masses 
larger than $\sim100$ M$_\odot$. Then we present a model for the origin
of this turn down that is consistent with the prevailing
picture of interstellar clouds, i.e., one in which pre-stellar clouds
have scale-free spatial structures and dynamical evolution on all scales. 

\section{A Statistical Limit for the Upper Stellar Mass}
\label{sect:stat}

For a universal IMF that is randomly sampled in a particular region over
time, the largest stellar mass that is expected to be present is given
by the expression $\int_{M_{max}}^{\infty}n(M)dM=1$, while the total
cluster mass in the
power law part of the IMF is
\begin{equation}M_{cluster}=\int_{M_{min}}^{\infty}Mn(M)dM=
{{xM_{max}^xM_{min}^{\left(1-x\right)}} \over{x-1}}, 
\label{eq:cluster}\end{equation}
given a power law slope $-1-x$ in the expression
$n(M)dM\propto M^{-1-x}dM$. For the Salpeter slope of $x=1.35$ and a
lower limit to the power law part of $0.3$ M$_\odot$, this integral is
\begin{equation}M_{cluster}\sim 3\times10^3
\left({{M_{max}}\over{100\;{\rm M}_\odot}}\right)^{1.35} {\rm
M}_\odot\label{eq:maxmass}.\end{equation} 
This equation does not apply to clusters that are old enough to have
lost their most massive members via supernovae; then the maximum current
stellar mass will be lower than the initial value.  It also does not
apply to the low-mass flattened part of the IMF. It applies only
to the stars in the power-law part of the IMF that 
are less massive than the most massive surviving member of the contemporary
cluster.  In fact, young star clusters always have masses that
are just large enough to account for their largest stars.  That is,
there are usually only a few stars, or perhaps just one, with a mass
close to the maximum stellar mass in the cluster.  Violation of equation
(\ref{eq:maxmass}) would require a significant number of stars in the
highest-mass bin of the IMF, with a sudden lack of stars any more massive
than this.  Without such a violation, there is no evidence for the existence of
an absolute largest stellar mass.

The situation changes when all of the current star formation
in a galaxy is considered as the statistical ensemble. 
According to our model, stellar masses are
randomly chosen from the same IMF regardless of where they form, 
so the IMF from each star-forming region and the summed IMF from all
star-forming regions in a galaxy are the same.
This is in agreement with observations, which suggest 
that the IMF from integrated light in a galaxy (e.g.,
Bresolin \& Kennicutt 1997) is the same as the IMF in individual
clusters
(see review in Elmegreen 1999b). When such random sampling
is considered in the context of a pervasive and
interconnected interstellar fractal
structure, then $M_{cluster}$ in equation (\ref{eq:maxmass})
can be very large. Indeed, the entire ensemble
of molecular clouds, along with their internal dense substructures and
peripheral atomic structures, is apparently a continuum of forms extending
over a wide range of cloud and intercloud media (Elmegreen \& Falgarone
1996; Elmegreen 1997a). For example, the interstellar media in
the Small Magellanic Cloud (Stanimirovic et al. 1999) and M81 group
galaxies (Westpfahl et al. 1999) look like continuous power law
structures with no characteristic scales in either the clouds or the
spaces between them.

What this continuity means is that the IMF could really be sampling from
a total gas mass that is the summed mass over many neighboring
star-forming clouds, instead of just the mass in any one GMC core. 
Neighboring clouds are just parts of the larger structure in
the overall interstellar hierarchy. This
summed mass might be more like $10^6$ M$_\odot$ of gas, corresponding to
$\sim10^{4.5}$ M$_\odot$ of stars considering typical efficiencies. Then
$M_{max}$ in the above expression would be $\sim600$ M$_\odot$.
The summed mass for sampling could even be the whole 
molecular interstellar medium. 
If the main sequence lifetime of a massive star is 2 My (Massey
1998), then typical galactic star formation rates of 5 M$_\odot$ per
year will have $10^7$ M$_\odot$ of young main sequence stars at any one
time. If these stars are considered as the statistical sample for
the IMF, then the maximum stellar mass would be 40000 M$_\odot$
for an extrapolated Salpeter function. 
Obviously the IMF has to turn down 
somewhere above $\sim100$ M$_\odot$ if
interstellar structures on kiloparsec scales provide
an ensemble for sampling the IMF.

These observations suggest that the power law IMF drops more
rapidly than the Salpeter slope above several hundred solar
masses, giving, in effect, two characteristic masses to the IMF, one at
each end. These two characteristic masses may have different origins and
scale differently with cloud properties, or they could be related and
scale together. The model proposed next is in the latter category. 
Other models are in Section \ref{sect:other}.

\section{A high-mass turndown in the random sampling model for the IMF}
\subsection{A timing limitation converted into a mass range}
\label{sect:timing}

The first stars that are likely to form in a cloud are those collapsing
from the densest pieces, which evolve the quickest. These are the lowest
mass pieces according to the turbulence scaling laws discovered by
Larson (1981), and, in the random sampling model, they give the smallest
stars. An important aspect of interstellar cloud structure is that these
low mass clumps are themselves clumped together into larger cloud
fragments, which, in turn, are clumped further into even larger pieces
(see reviews in Scalo 1985; Elmegreen \& Efremov 2000).
Such hierarchical structure gives the fractal aspect of interstellar
clouds that has been discussed extensively in the literature. 
Fractal structure has been found at the edges of atomic
(Vogelaar \& Wakker 1994), molecular (Dickman, Horvath, \& Margulis 1990;
Falgarone, Phillips, \& Walker 1991), and dust (Beech 1987; Bazell \&
D\'esert 1988; Scalo 1990) clouds, in the intensity distributions across
clouds (Stutzki et al. 1998), and in the clump-to-clump, and
cloud-to-cloud distribution of masses, considering a hierarchy of
structures that extends up to galactic scales (Elmegreen \& Falgarone
1996). 

In the random sampling model, stars can form from any level in this
hierarchy of structures as long as the gas can be made significantly
self-gravitating by any of a variety of physical processes.  In fact,
only a small range of hierarchical levels is necessary to give the
observed factor of $\sim100$ for stellar mass in the power law part of
the IMF. This mass factor corresponds to a size factor of $\sim7$ for
pre-stellar clumps, considering the scaling between mass $M$ and size $S$
that comes from the fractal dimension $D\sim2.3$: 
$M\propto S^{2.3}$ (Elmegreen \&
Falgarone 1996; Heithausen et al. 1998). Thus all stars in the power law
part of the IMF start within a limited range of scales in the interstellar
fractal, from the smallest gravitationally unstable piece at several
tenths of a solar mass up to fragments only $\sim7$ times larger in size. 
The whole
interstellar
fractal probably spans a factor of $10^5$ or more in size, although it
need not be continuous everywhere.
This limited range for all star formation, out of the
enormous range of available cloud structures, emphasizes the need for a
physical limitation to the mass of the most massive star, as there is
a physical limitation to the lowest-mass star.

Here we consider
what happens if small stars form so much faster than the evolution time
of the larger clump that contains them that the cloudy gas around the
smallest stars has time to re-establish a fresh hierarchy of structures
before the larger clump gets a chance to form its larger star. This new
structure might be driven by a variety of processes, including
small-star winds and continued
turbulence decay. Whatever the cause of mixing, the large-scale
gaseous structure that was available at first for the formation of a
large star will get redistributed into smaller pieces and end up making
only smaller stars. Thus, the large star will have a reduced
chance of forming. In terms of actual timescales, this constraint
corresponds to a formation time of the largest star that has to be
within a factor of only a few times the formation time of the smallest
stars in any one branch of the hierarchical tree of gas structures. This
factor follows from the relatively short timescale for turbulence to
redistribute and reform gaseous structures inside each existing scale,
and from the relatively short time for young stellar winds to stir and
distort the surrounding gas. 

Considering the turbulent scaling relationships for molecular clouds,
this constraint on timing can be translated into a constraint on mass.
The turbulence correlations show that the total linewidth or velocity
dispersion in a clump, $\Delta v$, scales with the clump size, $S$, to
about the 0.4 or 0.5 power (Larson 1981; for a compilation of clump
studies, see Efremov \& Elmegreen 1998). This means the crossing time,
$t_{crossing}\propto S/\Delta v\propto S^{\alpha}$ for $\alpha\sim0.5-0.6$.
The same clumps also have a relation between mass $M$ and size, which is
$M\propto S^\kappa$ for $\kappa\sim$ 2 to 3. In individual surveys,
$\kappa\ge3$ (e.g., Rosette: $\kappa=3.20\pm0.43$ -- Williams et al.
1994; Ophiuchus: $3.67\pm0.71$ -- Loren 1989; Maddalena-Thaddeus cloud:
$3.12\pm0.23$ -- Williams et al. 1994; M17: $3.18\pm0.56$ -- Stutzki \&
G\"usten 1990), while for the whole inner Galaxy, $\kappa\sim
2.38\pm0.09$ from
Solomon et al. (1987) and for all clouds taken together, $\kappa\sim2.3$
(Elmegreen \& Falgarone 1996). This latter value has been interpreted as
the fractal dimension by Pfenniger \& Combes (1994), Larson (1994),
Elmegreen \& Falgarone (1996) and Heithausen et al. (1998). 
In Larson
(1981), $\kappa\sim2$, which follows for virialized clouds with
$\alpha\sim0.5$. 
From these $\alpha$ and $\kappa$, the
crossing time scales with mass as \begin{equation}t_{\rm
crossing}\propto M^{\alpha/\kappa} \;\;\;{\rm for}\;\;\; \alpha/\kappa
\sim0.17-0.3.\label{eq:alpkap}\end{equation} 

In the examples below, we consider $\alpha/\kappa=0.2$ and 0.3. For
$\alpha/\kappa=0.2$, a constraint on the ratio of crossing times between
the smallest star-forming scale and the largest corresponds to a range in
stellar mass equal to this timing ratio to the power $\kappa/\alpha\sim5$.
Thus a factor  of $\sim100$ in stellar mass corresponds to a factor of
only $\sim2.5$ in turbulent crossing time. Regardless of the theory of
star formation, {\it most of the power-law range of the IMF is created in
regions with dynamical and turbulent time scales that are within a factor
of $\sim2.5$ of the time scale at the minimum gravitationally unstable
mass.}  This extremely tight schedule for the relative formation times
of intermediate and high-mass stars suggests that turbulence might be
involved, since turbulent fluids mix and destroy structures on similar
time scales. 

\subsection{A Stochastic IMF Model with a Timing Limitation}
\label{sect:model}

To illustrate the effects of timing constraints on the IMF, we modified
the previous models of the IMF that were based on root-density-weighted,
random samples of mass in hierarchically structured clouds (Papers
I--III). We used the formalism of Paper III to get the flattening at low
mass, which assumes there is a separate and uniform mass distribution
for the formation of stars inside each clump.

We simulate the dynamical competition for gas
mass in evolving turbulent clouds by introducing the additional
probability 
\begin{equation}
P(M)=e^{-\omega t(M)/t(M_J)}\sim e^{-\omega \left(M/M_J\right)^{\alpha/\kappa}}
\label{eq:exp}
\end{equation}
for the formation of a star of mass $M$, where $t(M)$ is the dynamical
time scale for regions of mass $M$ and $\omega$ is
a dimensionless parameter of order unity.  

There are several
ways to view this exponential in physical terms. 
If low mass stars mix up the gaseous structures in their
immediate neighborhoods, triggering other low mass stars
directly or in some other way preventing the gas from 
coming together into a single massive star, 
then the timescale for this mixing is on the order of a few crossing times
at the low-mass end. This time includes the star formation time for the
low mass stars, as well as the mixing time itself, assuming
the star formation time is relatively quick in terms of the local
dynamical time (Elmegreen 2000b).
If turbulence alone mixes the gas as part of the continued
decay of turbulent energy by shocking, viscosity, and magnetic diffusion,
then the time during which the massive structures are free to evolve
into massive stars is limited again to a few crossing times
(Stone, Ostriker, \& Gammie 1998; MacLow, et al. 1998).
Thus there is a window of opportunity to form a star of mass $M$ in
gas that also forms other stars, and this window has a
characteristic time scale equal to the crossing time on the
lowest mass scale, $M_J$, multipled by some factor of order unity,
$\omega^{-1}$.  

From a mathematical point of view, the exponential is the
Poisson probability that no significant mass redistribution inside scale
$M$ has occurred, given a mean number of redistribution times equal to
$\omega t(M)/t(M_J)$. Recall that the Poisson probability for $N$ events
given an expected number of events $\lambda$ is $\lambda^N
e^{-\lambda}/N!$, so the case of no events, $N=0$, has probability
$e^{-\lambda}$. Here, the waiting time for redistribution is taken equal
to $\omega^{-1}$ times the crossing time at the lower mass end, which is
the Jeans mass. If $\omega=1$, then the gas structure at scale
$M_J$ completely redistributes itself in one crossing time on that
scale.

The thermal Jeans mass in equation (\ref{eq:exp}) is only one of
several interpretations for the low mass limit of star formation.
Another is that pre-main sequence winds, possibly induced by Deuterium
burning, set the lower limit to what can accrete onto a star (Nakano,
Hasegawa, \& Norman 1995; Adams \& Fatuzzo 1996). For the wind model,
the lower mass objects are assumed to continue accreting until they
reach at least this limit. The nature of the lower mass limit is not
important for the present paper. Perhaps there is a combination of
effects, including both the requirement for clump self-gravity (the
$M_J$ limit) and the self-limitation by winds. In any case, we use the
notation $M_J$ to represent the lower limit to the power law part of the
IMF, where it turns over to a somewhat flattened distribution at lower
mass. 

In the algorithm for determining the IMF, the probability $P$ in
equation (\ref{eq:exp}) is applied after the clump mass is chosen from the
hierarchical tree. The whole algorithm runs as follows: A hierarchical
tree of masses inside other masses is set up initially using a random
number generator, with $H=10$ hierarchical levels and an average of
$N=3.2$ subclumps per clump, distributed as a Poisson variable in the
interval from 1 to 5 (this choice of $N$ is discussed below). These
masses are then chosen sequentially, using more random numbers, with a
weighting for clump choice that scales with the square root of the local
density in the ``cloud.'' Density is related to mass using the fractal
dimension $D=2.3$. This sequential choice gives a realistic aspect of
competition for mass because any clump that is removed to make a star
cannot be used again later to make a different star at a higher level.
The square root of density enters because that is the relative rate for
most of the dynamical processes that make stars. After this one mass is
chosen, we generate another random number, $R$, uniformly distributed
between 0 and 1, and keep that mass for
the final IMF if $P(M)>R$, where $P$ is given by equation (\ref{eq:exp})
with $\omega=1$, $\alpha/\kappa=0.2$, and $M_J=2$ in program units. This
simulates the second aspect of competition for mass, namely the
requirement that large regions not be given much time to turn their gas
into massive stars, considering that the smaller regions inside of them
should evolve more quickly. To get the actual star mass, we take a
random fraction of this chosen clump mass. This random fraction is
sampled from a probability distribution function that is flat in
logarithmic mass intervals, as discussed in Paper III; it gives a
realistic flat part to the IMF at low mass without affecting
the high mass power law. 

There are five parameters in the model, $M_J$, ${\cal R}$, 
$\omega$, $\alpha/\kappa$, and $D$.
The mass $M_J$ is the basic scale that determines whether or
not stars will form in a cloud; in the model it is the 
lower limit to the power law part of the IMF and so is
directly observable. The dimensionless
parameter ${\cal R}$ is the ratio of the largest to the smallest mass
for stars that form in any one clump; it should
be the same as the mass range for the flat part of the IMF and 
is therefore also observable. The
dimensionless quantity $\omega$ is the rate of significant mixing
at the scale $M_J$ 
multiplied by the dynamical time on that scale; 
a high mixing rate makes it difficult
to form massive stars, so $\omega$ affects where the high mass turn down
occurs. The ratio $\alpha/\kappa$ affects the conversion of mixing
time to mass; this depends on the scaling properties of the
turbulent fluid as discussed in Section \ref{sect:timing} and is
therefore observable, although inaccurately known. 

The fractal dimension $D$ enters only
into the conversion of mass to density (and therefore the relative sampling
rate).  The power law slope in the model does not
depend much on any of these parameters; it depends
mostly on $D$, but only by the amount $0.5-1.5/D$, which has a relatively small
range for $D$ between 2 and 3.  

The other parameters affect the parts
of the IMF that are beyond the power law, at low and high masses.
We have remarked previously how the slope of the IMF should not vary
much from region to region because it depends on the geometric
properties of the cloud, such as $D$. The endpoints of the power law
part of the IMF should depend on the details and physical parameters of
star formation, however, and that is the case for the dependence of the
present models on the parameters $M_J$, ${\cal R}$, $\omega$, and
$\alpha/\kappa$. 

Figure 1 shows the results of two numerical simulations in randomly
generated hierarchical clouds having $H=10$ levels with an average of
3.2 sub-clumps per level. The bottom panel is the
IMF and the top panel is the IMF multiplied by $M$, giving the relative
mass distribution rather than the relative number distribution. In both
panels, the dotted line is the new result, including the exponential
probability to simulate dynamical competition from equation
(\ref{eq:exp}), and the solid line is without this exponential.
The stellar masses in the dimensionless units used by the
computer program are given along the top axes of each panel. In these
units, $M_J=2$. The masses in physical units are given along the bottom
axes. For physical units, we choose $M_J=0.3$ M$_\odot$ from the
definition of the thermal Jeans mass (cf. Papers I-III).

The $P=0$ case for the solid line in figure 1 was run in order to
normalize the desired result against computer limitations. The main
limitation is the computer memory (2 GBy per processor), which would
cause a turn down at high mass if $N^H$ is small because of an inability
for the hierarchical tree in the model to span the observed stellar mass
range (see Paper I). For these runs, $N^H$ is sufficiently high that
there is no artificial 
turn down. The normalization run also indicates how much
noise should be present in the result. To minimize noise, we ran this
model for a large number of randomly generated trees until the number of
final stars equaled $\sim 2\times10^6$. This is comparable to the
total number of stars in the Milky Way that are younger than $\sim0.2$
My, if the birthrate is $\sim10$ stars per year. The stellar masses are
binned in intervals of 0.05 in the logarithm (base 10), and the numbers
on the ordinate of the lower panel are the numbers of model stars
actually chosen for these bins. Because this number is low for the
massive stars, the noise is relatively large there. Note that we could
not have limited
this simulation to high mass stars to get better accuracy there; a
competition for mass among all of the levels in the hierarchy is
necessary to get the IMF.

The expected mass range for the calculation can be determined from the
details of the algorithm. The smallest masses in the model are the
building blocks for all of the clumps, and they occupy the range from
0.01 to 1 program units, which are the units indicated at the top of the
figure. These smallest masses are distributed as $M^{-1}d\log M$ from
0.1 to 1 to simulate a continuing hierarchical structure below $M_J$
(see Paper I). Thus the average mass of the smallest unit is 0.2558
program units. The choice of star mass from a given clump mass
introduces another factor. This comes from a distribution that is
uniform in $\log \epsilon$ for $\epsilon$ equal to the star/clump mass
fraction. For a total range of $\epsilon={\cal R}=30$, as assumed here,
the average value of $\epsilon=0.2842.$ The average number of subclumps
is chosen to be 3.8, but computer memory limitations force us to allow
variations in the actual number of subclumps per clump that only range
from 1 to 5. The average of a Poisson variable in the interval from 1 to
5 with a mean of 3.8 equals 3.2, which is the effective average number
of subclumps per clump in the simulations. With these three numbers in
runs with 9 hierarchical sub-levels (the tenth level is the whole
cloud), the model IMF is expected to begin to drop significantly because
of computer limitations at a mass equal to $M_J$ times
$0.2558\times0.2842\times3.2^9=2560$. This is larger than the maximum
masses for the lines in figure 1 (in terms of $M_J$), so there are no
significant drops in the IMF from memory limitations.

A comparison between the two lines in figure 1 illustrates the main
point of this paper: the timing constraint steepens the IMF a little for
the whole power law range, actually making the slope closer the Salpeter
value, and it produces a noticeable (factor of 10) dip in the IMF below
the $P=0$ case at masses larger than $\sim1000 M_J$ ($=300$ M$_\odot$ in
the figure).

For a realistic physical model, the mass at the high mass turn-down
should depend on the average number of crossing times for the
redistribution of mass inside each particular mass scale. We have
assumed in the above calculations that only one crossing time
($\omega=1$) is enough to completely mix up a clump and render it
improbable to convert a high fraction of its mass into a single star.
This may be too long a time. If it takes $1/2$ crossing time for mixing
to have this effect, then we should have written
$P(M)=e^{-2\left(M/M_J\right)^{\alpha/\kappa}}$ in equation
(\ref{eq:exp}). This has the effect of decreasing the mass at the upper
mass turn down, which depends sensitively on this factor in the
exponent. We consider this change for figure 2, discussed below. 

\section{Upper Stellar Mass Cutoffs}
\label{sect:other}

The high mass turn down in the model IMF shown by figure 1 is enough to
account for the lack of supermassive stars in most star-forming regions,
and it agrees well with the observation of a Salpeter slope out to
$100-130$ M$_\odot$ in the R136 cluster (Massey et al. 1995b; Selman et
al. 1999). It still predicts too large a value for the maximum stellar
mass in a whole galaxy disk, however. 

Figure 2 shows IMFs and the expected maximum initial 
stellar masses versus the
total stellar masses for four
cases: (1) a pure power-law Salpeter function above 0.3 M$_\odot$,
having the form $M^{-1.35}d\log M$, (2) an IMF of the form
$M^{-1.25}e^{-\omega\left(M/0.3\;{\rm M}_\odot\right)^{0.2}}d\log M$ for
$\omega=1$ that matches the Salpeter IMF at low mass and has the
proposed drop off at high mass from timing
limitations, as in the model of section \ref{sect:model}, (3) the same
as (2) but with $\omega=2$, and (4) the same as (2) but with a steeper
power law in the exponent ($\alpha/\kappa=0.3$). The pure Salpeter
function reproduces equation (\ref{eq:exp}), and the modified functions
indicate how much more stellar mass must be sampled to get a massive
star when there is a timing constraint.

Figure 2 suggests that if $\omega$ and $\alpha/\kappa$ are large enough
to avoid stars more massive than $\sim300$ M$_\odot$ in a whole galaxy
disk with $10^7$ M$_\odot$ of young stars, then the IMF slope out to 130
M$_\odot$ is too steep to explain the observations of R136. This result
is fundamental to IMF theory and not particularly dependent on the
present models: {\it There is a problem getting both the Salpeter
function out to $\sim130$ M$_\odot$ in dense clusters} (e.g., Selman et al. 1999)
{\it and at the same
time not getting any $\sim300$ M$_\odot$ stars at all in a whole galaxy.
} Some type of IMF cutoff at high mass has to begin just beyond the
observed mass for the most massive stars, and it has to be steep to 
avoid forming overly massive stars in whole galaxies. 

We can see six ways around this problem. First, supermassive stars really
could exist, particularly in other galaxies, but not be recognized or
resolved yet.  We discuss in Section \ref{sect:mj} how their presence
in starburst regions could support the IMF model.

Second, there could be a self-limitation of newborn stellar mass from
winds or radiation pressure beyond the mass of the largest observed
star. Self-limitation of stellar mass is reasonable but there are no
direct observations of it in the pre main-sequence phase and theoretical
considerations for how it might come about are beyond the scope of
this paper.

Third, supermassive stars could form with the relative abundance
predicted by the modified IMF models, but then erode so quickly during
the first few hundred thousand years of the main sequence that they do
not come out of their primordial cloud cores at their full initial
masses. In this case, they would be observed primarily as embedded
ultra-luminous infrared sources and not be seen otherwise. 

Fourth, there could be a limit to the mass of a cloud in which coherent
star formation samples the complete IMF. The upper mass limit for a
cloud that is necessary to avoid clusters much larger than $10^{5.5}$
M$_\odot$ (at which point the maximum stellar mass begins to exceed the
maximum observed mass for the $\omega\sim1$, $\alpha/\kappa=0.2$ models)
is in fact the upper mass limit that is really observed for clouds,
namely $\sim10^7$ M$_\odot$ for the giant spiral arm complexes
(Elmegreen \& Elmegreen 1983, 1987; Rand 1993, 1995), considering an
efficiency of $\sim3$\%. Since most stars form in these complexes
anyway, we only have to assume that each one contains an independent
hierarchical tree of cloud structure and that IMF sampling does not span
the distance between them. The same would be true for the giant
cloud complex in the LMC that formed the 30 Dor region, perhaps with
a slightly higher efficiency. 

In a small galaxy like the LMC, this sampling limit can explain the
maximum stellar mass without recourse to timing or other limitations
during the star formation process. In massive galaxies, however, the
analogous argument becomes more questionable. If there are many $10^7$
M$_\odot$ clouds in a galaxy, then the combined mass available for star
formation can be much larger than $10^7$ M$_\odot$, and in this combined
mass, purely random star formation would sample the IMF out to
prohibitively large stellar masses. To get around this problem, we would
have to assume that several independent $10^7$ M$_\odot$ clouds do not 
sample the IMF in the same way as one extended cloud (or spiral
arm) with the same total mass. But if we make this assumption,
we have lost one important advantage of the random sampling model, i.e.,
that star formation on the sub-parsec scale does not have to know
about gas on the kiloparsec scale. If $10^7$ M$_\odot$ is
really a limit to the horizon of random sampling, then a
$4\times 10^3$ M$_\odot$ clump inside such a cloud, which might try to
make a $10^3$ M$_\odot$ star all at once if it were in a bigger cloud, 
has to know instead that there are no other
$10^7$ M$_\odot$ clouds immediately nearby that would
increase the total gas mass available for sampling. We view such
environmental knowledge as unreasonable. If star formation processes are
really local, then the mass of the peripheral cloud cannot affect the
IMF except to influence the total number of stars that are formed. 

Fifth, there could be a bias in the sampling of star mass as a function
of cloud mass that is not from the formation of stars, which was
considered to be unlikely in the previous paragraph, but from the
destruction of clouds. It is possible that stars form randomly
everywhere with no local knowledge of the overall cloud mass around
them, but that star formation stops in a cloud of a certain mass once a
star of a certain mass forms. Since there are more low mass clouds than
high mass clouds, an increasing function of maximum star mass versus
cloud mass at the time of destruction will steepen the IMF (Paper II).
This is because low mass clouds are more easily disrupted than high mass
clouds, and as a result, are able to form stars only up to a certain low
mass compared to the stars that are able to form in high mass clouds.
With such a bias, the larger number of low mass clouds ensures there
will be more star formation episodes with a small maximum mass than with
a large maximum mass. We discussed previously (Paper II) how this
process could explain the apparently steep IMF in the extreme field
regions of the LMC and Galaxy (Massey et al. 1995a), but concluded 
that the similarity between the cluster IMF and the whole-galaxy IMF
limited this steepening effect to only a small fraction of the total
mass. The onset of rapid and thorough cloud destruction above $\sim10^5$
M$_\odot$ of gas could in principle steepen the IMF for stars greater
100 $M_\odot$ without any of the timing considerations modeled in the
previous section.

A sixth possibility is that there are different IMFs in different regions,
with shallow slopes in dense clusters to satisfy the constraint from
R136, and steep slopes in other regions to give a low maximum mass. The
constraint on these parameters from R136 is for an extremely dense,
self-gravitating core.  If most star formation is not in such cores but in
lower density regions, then $\omega$ and $\alpha/\kappa$ could be higher
in general, and supermassive stars could be avoided outside these cores.

In summary, the modified IMF with a timing constraint discussed
in the previous section explains the
observed IMF and the lack of $\sim300$ M$_\odot$ stars in normal
star-formation regions and clusters, which is something the pure
Salpeter function cannot do. However, no single model of the type
discussed here can explain the simultaneous appearance of a Salpeter IMF
out to $\sim130$ M$_\odot$ in R136 and a lack of $\sim300$ M$_\odot$
stars in a whole-galaxy sample. Either supermassive stars limit their
own mass or destroy their own clouds excessively compared to low mass
stars, or an excess of massive stars forms in dense clusters.
The next section considers this last possibility in more detail.

\section{An Excess of High Mass Stars in Clusters?}
\label{sect:coal}

A possible difference in the IMF for high and low density environments
was mentioned in the previous section as a way to explain the relatively
shallow IMF out to $100-130$ M$_\odot$ in dense clusters like R136 and
at the same time explain the relatively steep drop off beyond this mass
that is imposed by the lack of $\sim300$ M$_\odot$ stars in whole
galaxies. One thing that would do this is protostar coalescence or
enhanced protostar accretion in dense clusters but not elsewhere. Then
any tendency for the galaxy-wide IMF to decrease at high mass from timing
constraints or other causes can be compensated by a slight overabundance
of $50-130$ M$_\odot$ stars in dense clusters. At the moment, there is no
direct observation of protostar coalescence, and there are observations of
massive stars in the 30 Dor region that are not in the dense R136 core,
so perhaps this possibility is unreasonable. Nevertheless, we consider
the implications here briefly.

The observation of massive stars in extremely dense environments has often
led to the idea that some of the largest stars may form by coalescence
(Zinnecker 1986; Larson 1990; Price \& Podsiadlowski 1995; Bonnell et al.
1998; Stahler, Palla, \& Ho 2000) or enhanced accretion (Larson 1978;
Larson 1982; Zinnecker 1982; Bonnell et al. 1997).  Coalescence of
pre-main sequence stars requires extreme densities, even higher than what
is observed in the densest clusters, but coalescence after accretional
drag (Bonnell et al. 1998) or coalescence of protostars with extended
disks (McDonald \& Clarke 1995) may be possible.  We showed in Paper
II that even protostellar coalescence is likely to operate at a rate
that scales with the square root of the local density, so the same basic
model of the IMF should apply if stars routinely coalesce. The question
is whether high mass stars form more by coalescence than low mass stars,
because if they do, then the self-similarity assumed by the model would
not be appropriate.

There is considerable evidence that the IMF is steeper in regions of
lower density, but it is unclear how much of this evidence is free from
selection and systematic effects.  For example, local field-star IMFs,
corrected for past birthrates and vertical Galaxy drift, typically give
$x$ in the range of 1.5 to 2, instead of $\sim1.35$ (Miller \& Scalo 1979;
Garmany, Conti, \& Chiosi 1982; Humphreys \& McElroy 1984; Scalo 1986;
Blaha \& Humphreys 1989; Basu \& Rana 1992; Kroupa, Tout, \& Gilmore
1993; Parker et al. 1998).  The same steep IMF has been derived by Brown
(1998) for local OB associations using Hipparcos positions for membership.
LMC clusters in regions of low young-star density (J.K. Hill et al. 1994;
R.S. Hill et al. 1995) and unclustered embedded stars in Orion (Ali \&
DePoy 1995) have $x\sim1.5-2$ also. An extreme example is the steep
high-mass IMF ($x\sim4$) in the field regions of the LMC and Solar
neighborhood (Massey et al. 1995a).

These relatively steep IMFs contrast to the shallow IMFs often found in
dense clusters and in normal cluster cores (Sagar et al. 1986; Sagar \&
Richtler 1991; Hunter et al. 1996a,b, 1997; Selman, et al. 1999; see
review in Massey 1998, and Fig. 5 in Scalo 1998).

The problem with these measurements is that systematic effects are
expected that would give such a density dependence
even if the IMF at birth is
the same on average.  Mass segregation could make the IMFs in cluster
cores shallower than at the edges, and differential drift between the
short-lived, high-mass stars and the long-lived, low-mass stars that are
leaving OB associations can overpopulate the field with low mass stars.
Unknown star formation histories in OB associations can also produce
uncertainties in the upper mass IMF when corrected for evolved stars that
are no longer present.  There has been little modeling to quantify these
effects, except for mass segregation, and that seems
too slow to produce the observed flatness in the youngest cluster cores
(e.g., Subramaniam, Sagar, \& Bhatt 1993; Hillenbrand \& Hartmann 1998;
Fischer et al. 1998; Bonnell \& Davies 1998).  If this is the case,
then massive stars would be born preferentially in the cores, perhaps
as a result of the coalescence or accretion mechanisms discussed above.

Unfortunately, no one knows
whether the IMFs at the faint edges of clusters are steep enough
to compensate for the flat IMFs in the cores, giving the same IMF on average
as in non-clustered regions.  Nor are the dynamics of a cluster during
star formation known well enough to rule out mass segregation.
For example, Giersz \& Heggie (1996) found that mass segregation is
virtually complete in only one core collapse time. This implies that if
magnetic pressure is comparable to turbulent pressure in a cloud core,
and so the initial motions of the stars are sub-virial, leading to a
significant and rapid collapse of stellar mass to the cloud core, then
by the time we see a tight cluster in the core, the segregation of high
and low stellar mass is finished. 

The observation of dense clustering around high mass stars has also
been used to suggest that massive stars require coalescence or excessive
accretion (Testi, Palla, \& Natta 1999).  However, the rarity of massive
stars compared to low mass stars implies that any region of star formation
will generally form a lot of low mass stars along with the few high
mass stars.  The correlation between the mass of the largest star and
the density of the cluster, found in their survey (Fig. 7), is also
a correlation between maximum star mass and cluster mass, considering
the cluster radius is about constant, as shown in their figure 2.
Then the maximum star mass should correlate with cluster density because
of equation (\ref{eq:cluster}). For example, in figure 7 of Testi, Palla,
\& Natta (1999), the cluster density varies from $\sim10^{3.3}$ pc$^{-3}$
for an O5 star to $\sim10^{2}$ pc$^{-3}$ for an A0 star. The masses of
these stars vary from 40 M$_\odot$ at O5 to 4 M$_\odot$ at A0 (Mihalas \&
Binney 1981).  Then the ratio of cluster densities, $10^{1.3}$, which is
also the ratio of cluster masses for a constant cluster radius, is equal
to the ratio of stellar masses, $10$, to a power equal to the Salpeter
slope, $x=1.35$, as predicted by equation (\ref{eq:cluster}).  Thus the
trend between cluster density and maximum stellar mass found by Testi,
Palla, \& Natta (1999) could be the same as equation (\ref{eq:cluster}). A
similar point was made by Bonnell \& Clarke (1999).

The Trapezium cluster is a good place to check the coalescence model. It
is extremely dense, containing $\sim5000$ stars pc$^{-3}$ (Prosser et
al. 1994) or more (McCaughrean \& Stauffer 1994), and it has several
massive stars with close companions. According to Hillenbrand (1997),
Hillenbrand et al. (1998) and Weigelt et al. (1999), the Trapezium
primary stars $\Theta^1$A,B,C, and D have masses of 20, 7, 45, and 17
M$_\odot$, respectively, and they have 2, 4, 0, and 1 close, lower-mass
companions.

The number of massive stars expected in the Trapezium cluster can be
determined from an IMF extrapolation of the low mass star count. Palla
\& Stahler (1999) fit the IMF for stars below 10 $M_\odot$ to the recent
average IMF derived by Scalo (1998), which has slopes of $-0.2$ between
0.1 M$_\odot$ and 1 M$_\odot$, $-1.7$ between 1 M$_\odot$ and 10
M$_\odot$, and $-1.3$ between 10 M$_\odot$ and 100 M$_\odot$,
considering logarithmic intervals of mass. For a total number of
Trapezium stars equal to 258 in the mass range from 10$^{-0.4}$
M$_\odot$ to $10^{0.8}$ M$_\odot$, the resulting IMFs for these mass
ranges are $380M^{-0.2}$, $380M^{-1.7}$ and $150M^{-1.3}$, respectively.
If we fit the 258 stars to a single slope of $-1.3$, then the result is
$100M^{-1.3}$ for all masses.

The integral of these IMFs over mass from 30 M$_\odot$ to 100 M$_\odot$
equals 1.1 for the Scalo (1998) fit and 0.7 for the constant slope of
$-1.3$. This implies that the presence of a single 45 M$_\odot$ star in
the Trapezium cluster is not unusual. Similarly the expected number of
stars between 10 and 20 M$_\odot$ is 3.4 for the Scalo fit and 2.2 for
the constant slope. These compare well with the observed number of 2.
Thus the massive stars in the Trapezium cluster were just as likely to
form there as anywhere else with a normal IMF, including a much more
dispersed cluster or association having the same total mass. There is no
need to postulate that the 45 M$_\odot$ star or any of the other massive
stars required the companions to be there or that they formed by
coalescence. 

In summary, the IMF slope at intermediate to high mass is about the same
in clusters and associations spanning a factor of $\sim200$ in density
(Hunter et al. 1997; Massey \& Hunter 1998; Luhman \& Rieke 1998). This
suggests that theories of the IMF should not rely on processes that
require high densities, including coalescence, protostar interactions,
or enhanced accretion, to give a slope in the range $x\sim1-1.5$.
The lowest density regions of star formation have no protostellar
coalescence, yet their intermediate-to-high mass IMFs look about the
same as in dense clusters, to within the statistical noise, and aside
from the expected effects of mass segregation, differential dispersal,
and post-main sequence evolution.  While this apparent IMF uniformity
would seem to make the theory of the IMF simpler, it is a problem
when it comes to understanding the upper mass limit. Why isn't the
maximum stellar mass larger in extended regions of star formation than
it is in the most massive dense clusters, where $\sim130$ M$_\odot$
stars are observed among only $\sim10^4$ $M_\odot$ of other stars?
The upper mass cutoff that is necessary seems to be sharper than what
is likely to arise from self-similar turbulence and timing constraints.
Perhaps self-limitation of maximum stellar mass is the only alternative.

\section{Variations in M$_J$}
\label{sect:mj}

If the proposed top-heavy IMF in starburst regions (Rieke et al. 1980)
is the result of a higher minimum mass, $M_J$ (Paper I), then,
without self-limitation, there
should be unusually massive stars in these regions too, according to the
IMF model presented here. 
Stars with initial masses of $\sim10^3$ M$_\odot$ at their
Eddington limits might be possible, forming from the larger clumps in
normal (but warm) clouds. The presence of such stars would have
important consequences, such as a higher luminosity-to-mass ratio for a
stellar population, a higher supernova or gamma ray burst rate per unit
gas mass, stronger winds and galactic outflows per unit gas mass, as
well as unusual abundance ratios from nuclear processing. The
supermassive stars themselves might look odd too, having unusual stellar
spectra indicating rapid mass loss, or possibly appearing primarily in
the infrared (Hoyle, Solomon \& Woolf 1973). The post-main sequence
stars could include the proposed ``warmers'' (Terlevich \& Melnick 1985)
or an unusual abundance of WR stars compared to other stars, as in WR
galaxies (e.g., Ohyama, Taniguchi, \& Terlevich 1997). {\it Any observation
of an IMF that is shifted entirely to larger masses, from the lower mass
turnover to an upper mass turndown, would support the IMF model
developed here.} 

The consequences of a systematic shift toward lower stellar masses
resulting from a smaller value of $M_J$ would be important too. This may
allow Brown Dwarfs to form without normal stars at ultra-low temperatures
(Elmegreen 1999c) or extremely high-pressures (Fabian 1994).

\section{Summary}

Turbulent and young-star mixing of star-forming gas can limit the
masses of the largest stars by continuously forming smaller-scale
structures and smaller stars inside all of the larger regions. A model
for this mixing uses a time constraint for
star formation at every mass $M$ that decreases exponentially with the
turbulent crossing time at that mass, normalized to the crossing time
at the lowest significant mass, which is taken to be $M_J$. The results
were presented in the form of a random sampling simulation in figure 1
and with the analytical form in figure 2.

The simulations suggest that the timing constraint makes the slope of
the IMF slightly steeper by eliminating a small fraction of the
intermediate mass stars compared to the low mass stars, but the greatest
effect is for the high mass stars, which are significantly reduced
compared to the Salpeter function. The result is a Salpeter IMF starting
at a lower mass limit of several $M_J$ and continuing to higher masses
with a gradual turn down in the upper mass range. At masses of about
$10^3$ M$_J$, the turn down amounts to a decrease by a factor of
$\sim10$.

These results explain the lack of young stars larger than several hundred
solar masses in giant star-forming regions or spiral arm pieces where
random sampling from the Salpeter function alone would predict several
such stars. The results do not explain how dense clusters like R136
can have a Salpeter function out to $\sim130$ M$_\odot$ while a whole
galaxy does not have stars greater than $300$ $M_\odot$. That is, the
required fall off at masses greater than $130$ M$_\odot$ is too steep
for the assumed timing constraint to explain. There are several ways
out of this problem, discussed in Section \ref{sect:other}, including
an excess of high mass stars in dense clusters, where the Salpeter IMF
is commonly measured, and self-limitation of the maximum star mass.

Acknowledgement: 
Helpful comments from the referee are gratefully acknowledged, as are
discussions with Kris Davidson about supermassive stars.

\newpage
\begin{figure}
\vspace{6.2in}
\includegraphics{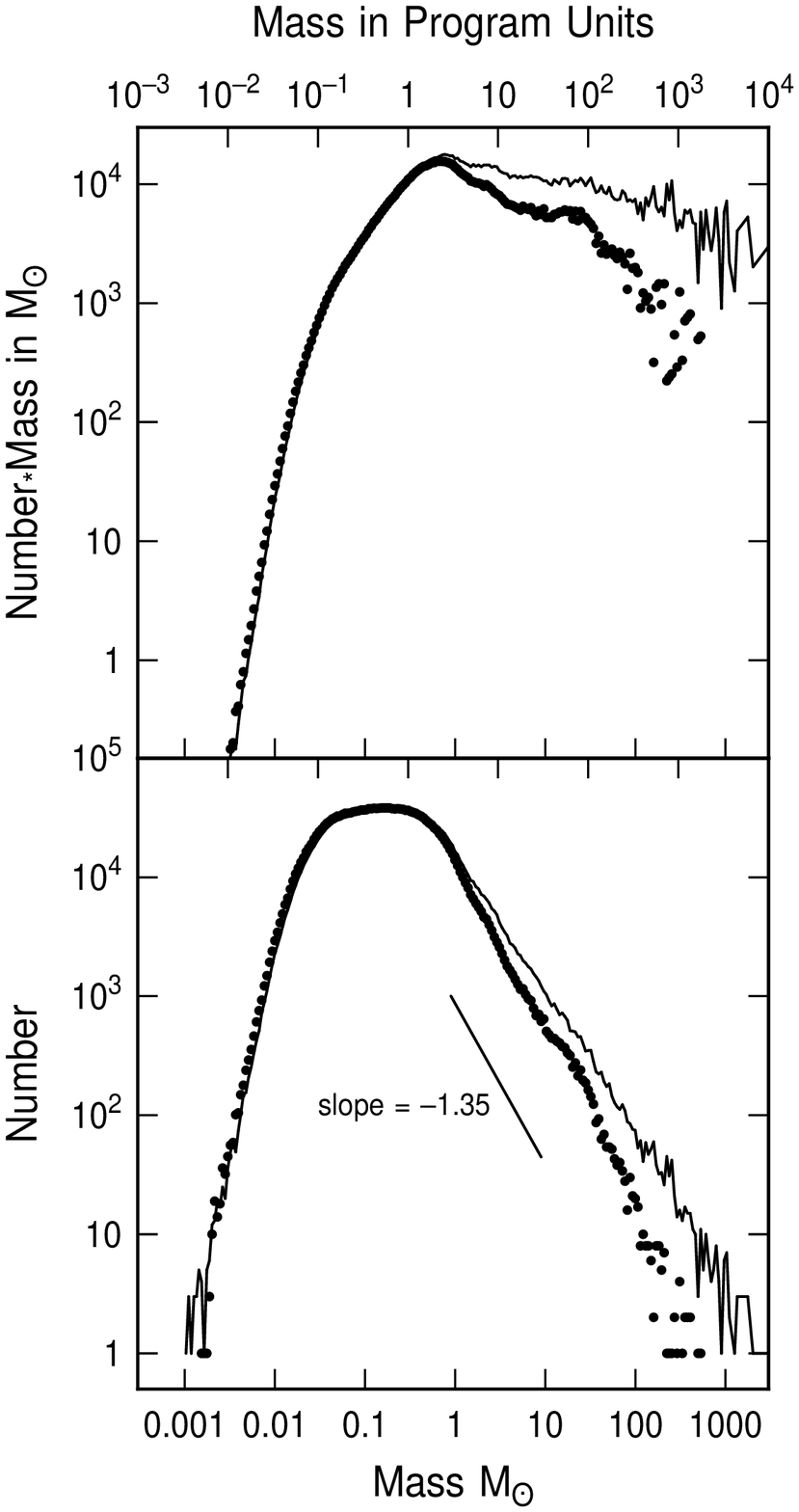}
\caption{The IMF simulation is shown in the bottom
panel for two cases: the solid line is for a model
with no timing constraint in the choice of
clumps for star formation, while the
dotted line is with the timing constraint, assuming
$\omega=1$ and
a power in equation 3 equal to 0.2.
Both models contain $2\times10^6$ stars.
The top panel shows the product of the mass and the IMF.}
\end{figure}

\newpage
\begin{figure}
\vspace{6.2in}
\includegraphics{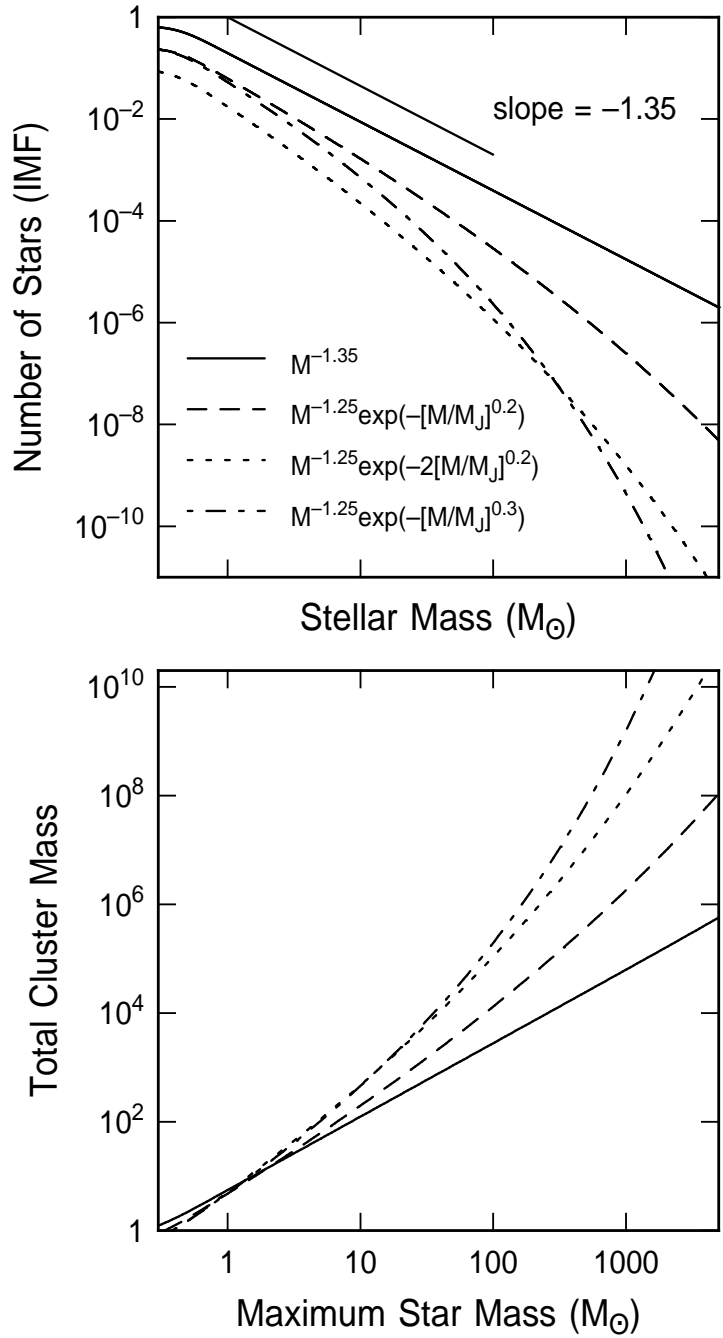}
\caption{IMF models versus stellar mass (top), 
and total cluster masses versus maximum
stellar mass.}
\end{figure}

\end{document}